\documentclass[12pt,letterpaper]{JHEP3}
\pdfoutput=1
\usepackage{cite}
\usepackage{epsfig}
\usepackage{subfigure}

\graphicspath{{Figures/}}

\title{Probability of Slowroll Inflation in the Multiverse}

\author{I-Sheng Yang\footnote{isheng.yang@gmail.com}\\
ISCAP and Physics Department \\
Columbia University, New York, NY, 10027 , U.S.A.}

\abstract{Slowroll after tunneling is a crucial step in one popular framework of the multiverse---false vacuum eternal inflation (FVEI).  In a landscape with a large number of fields, we provide a heuristic estimation for its probability.  We find that the chance to slowroll is exponentially suppressed, where the exponent comes from the number of fields.  However, the relative probability to have more e-foldings is only mildly suppressed as $N_e^{-\alpha} $ with $\alpha\sim3$.  Base on these two properties, we show that the FVEI picture is still self-consistent and may have a strong preference between different slowroll models.}

\begin{document}

\section{Introduction}

False vacuum eternal inflation\cite{GarVil97,Vil02,Vil04,GarSch05,GarVil05,GarGut06,SchVil06,
VanVil06,Vil06,BouFre06a,BouFre06b,Lin07,GarVil07,CliShe07,BouYan07,BouFre07,BouFre08a,BouFre08b,BouLei08,GarVil08,DGSV08,BouHal09,Bou09} is currently the most studied model of the multiverse scenario.  On the theoretical front, it is well motivated by the string landscape\cite{BP,DouKac06}.  On the observational front, if we see negative curvature from the Planck\cite{Planck1} data, it will become the only preinflationary cosmological model with real evidence\cite{FreKle05,KleSch12,GutYas12}.   

In this model, one or more vacua inflate forever and constantly decay into various other vacua.  Some of the decays lead to bubbles of open universes supporting our observed cosmology.  Since the dynamics involves exponentially small or large numbers, this picture can provide exponentially strong selection rules on which vacuum we live in.  One can check whether these rules are so strong that actually rule out this model by a direct contradiction with existing observations\cite{GarVil05,SchVil06,BouFre06b,Lin07,DGSV08,BouFre08b}.  When it does not, one can try to draw sharp predictions from these rules\cite{GarVil05,SchVil06,VanVil06,Lin07,CliShe07,BouYan07,BouFre08b,BouLei08,BouHal09}.

It is well-known that a bubble nucleation itself can only make an empty universe\cite{GutWei83}, which is incompatible with the rich structure we observed.  It will also make a curvature dominated open universe as opposed to the flat one we have.  The most well-accepted solution to these two problems is to have a period of slowroll inflation after the tunneling.  It will flatten out the curvature with enough number of e-foldings, generate perturbations to seed structure, and reheat into thermal particles of the big-bang cosmology.  However, there is no reason why after a vacuum decay, the state of a bubble universe will go through slowroll inflation.  Quantifying the probability for that to happen can give us further selection rules.  Implementing these new selection rules requires us to recheck for contradictions.  For example the rareness of slowroll inflation favors Boltzmann Brains\cite{DysKle02,Pag06,BouFre06b}, since they can be produced in a bubble without slowroll.  Also if longer inflation is strongly disfavored, it will be in conflict with our seemingly ever improving curvature bound.

Using the simplest toy model---standard gradient flow inflation with $N$ canonical scalar fields, we provide a heuristic estimation for the probability of slowroll inflation in Sec.\ref{sec-probability}.  In Sec.\ref{sec-global} we show that the regions supporting at least a few e-foldings of slowroll are exponentially rare, where the exponent comes from $N$, the number of fields.  On the other hand, in Sec.\ref{sec-efold} the relative probability to have more e-foldings is shown to be only suppressed by $N_e^{-\alpha}$, where $\alpha$ remains to be a small number even when $N$ is large.  Although these results mostly agree with numerical searches up to $N=6$\cite{AgaBea11}, in Sec.\ref{sec-attract} we further include the attractor effect, a multi-field effect which has not be explicitly addressed before, and confirm their validity at least in the large $N_e$ limit.  In Sec.\ref{sec-check} we check these results against the concerns mentioned in the previous paragraph and find no contradictions.

The selection rules studied here potentially lead to useful predictions. If the probabilities of all realizations of slowroll inflation are exponentially suppressed, then the relative probability between two different models has a chance to be exponential, too.  This means some models are much more favored than others.  So, the multiverse picture may provide a new preference among slowroll models.  Just ask the simple question: {\bf which model is more likely to have made our universe?}\footnote{One can try to ask this question in more general ways. For example, including cosmological scenarios other than slowroll inflation\cite{Leh12}, or other UV completions that initiated slowroll inflation\cite{DowDut12}.  In this paper, we restrict our attention to the combination of slowroll and tunneling, which is most conservative---requires only widely accepted results from field theory and semiclassical gravity.}

Our estimation bases on the na\"ive assumption that where a tunneling path ends in the field space has no correlation with whether that point supports slowroll inflation.  This is entirely due to our ignorance.  Currently there is not enough understanding on multifield tunneling paths to analyze such correlation.  In Sec.\ref{sec-dis} we suggest the sharpest possible manifestation of the multiverse selection rule for slowroll models---if certain slowroll model strongly correlates with the local property where a tunneling path ends, then it does not suffer from the exponential suppression.  Thus, if such model exists, it is the most likely slowroll inflation realized in the multiverse.  We point out a few recent works on multifield tunneling paths that may help us to search for slowroll models with this property.

\section{Probability for Multifield Slowroll Inflation}
\label{sec-probability}

Our heuristic estimation bases a model of $N$ scalar fields with a flat field-space metric.  The equations of motion in a homogeneous background are
\begin{eqnarray}
\ddot\phi_i + \frac{\dot a}{a} \dot\phi_i 
&=& -\frac{\partial V}{\partial \phi_i}~, \label{eq-field} \\
\left(\frac{\dot a}{a}\right)^2 &=& 
\frac{1}{3M_p^2}\left(\frac{\dot\phi_i^2}{2}+V\right)+\frac{1}{a^2}~.
\label{eq-GR}
\end{eqnarray}
Here the curvature is chosen to be negative, as we are interested in the cosmology of a bubble universe.  The initial condition is set by the analytical continuation of the tunneling Euclidean instanton, $\dot\phi=a=0$.  This is known as the ``open inflation''\cite{KamRat94,BucGol94,AmeBac95,Gar95,YamSas96,SasTan96,VilWin96,WhiSil96,AmeBac96,Lin98a,LinSas99,ChiYam99,BarAlb04,YamLin11,SugYam11,VauWes12}.  The instanton also contains the information about the tunneling path that begins near the parent vacuum.  The end point of that tunneling path sets the initial condition for $\phi$.

It is well-known that slowroll inflation has the overshoot problem\cite{BruSte92}---an order one $\dot\phi$ will rush through the region in the field space that is tuned flat to support slowroll.  The open inflation scenario ameliorates such problem\cite{FreKle05}.  Because $\dot{a}/a$ starts as infinity in Eq.~(\ref{eq-field}), $\dot\phi$ cannot easily reach order one even when the potential is steep.  If the potential is tuned to support slowroll in certain field space region $A$, there will be a corresponding ``attraction'' region $B\supset A$.  Potential is generally steep in $B$, but the field will still roll down to $A$ before it can acquire an order one velocity.  

Although it is possible to start outside $B$, acquire an order one velocity, and hope that it happens to slow down and enter a slowroll region at the same time, we think those cases are relatively less likely. We will focus on the probability that a tunneling path ends in an attraction region.  

In most of the works on slowroll inflation, one would separate the heavy and light fields and focus only on the ``low energy'' effective motion.  When a tunneling is involved, some may assume that it also involves the light fields only.  We should emphasize that such assumption is quite inappropriate for our purpose.  Tunneling is a non-perturbative process.  Whether a perturbative excitation in one direction is heavy or light has little to do with the possible tunneling coming that way.  With our currently limited understand on multifield tunneling paths, we should not have any preference.  We will simply assume that the end points have a uniform distribution per unit field space volume.  Since the typically quoted number of $10^N$ vacua on the landscape comes from at least $N$ dynamical fields, our focus will be on the effect of this large $N$.  We will estimate the fraction of $N$ dimensional field space volume in the attraction regions.  We shall do this in two steps.  First we estimate how popular the slowroll regions are.  Then we will weight each slowroll region with the size of the corresponding attraction region.

\subsection{Exponential Suppression of Slowroll Regions}
\label{sec-global}
In principle, we would like to search over the entire landscape and count the number of regions supporting slowroll inflation.  In practice, we need to turn the problem from a global search into a search over ensembles of random potentials, similar to many recent works\cite{AazEas05,LinWes07,TyeXu08,AgaBea11,KimLid11,DiaFra12,BatBat12,McARen12,BlaGom12}.  Note that ``random'' here is still a vague term.  For each parameter that takes a random value, one still needs to choose an appropriate weight function.  Currently no one knows the fundamental guideline for such choice, so strictly speaking it is arbitrary.  However, our goal is just a heuristic estimation focusing on the role of the number of fields, $N$.  Assuming no field is special from the global point of view, whether a quantity acquires the power $N$ is an unambiguous property independent of the weight function.

We treat $V$ as the final effective potential with all corrections taken into account, and applies no more restrictions other than it being a continuous and smooth function.\footnote{This is different from the majority of works on inflationary model building.  People quite often focus on potentials with special properties like slightly broken symmetries.  From our point of view, doing so is equivalent to ``zooming in'' on special sectors of the landscape, but the corresponding suppression factor on the probability is usually hard to quantify.  Our unrestricted and untuned $V$ is more appropriate to represent the global behavior of the landscape.  A slowroll-friendly region appears by accident---when the random combination of coefficients happens to be right.  We can then quantify the probability for such combinations.  Maybe one can take a closer look at each accidentally slowroll-friendly region and observe an emergent symmetry, but that is quite parallel to our purpose.}  Technically, our definition of ``supporting slowroll'' is more specific than necessary.  We consider only regions supporting the classical slowroll motion along the gradient flow.  Namely we focus on the situations where the $\ddot{\phi}_i$ term can be ignored in Eq.~(\ref{eq-field}) and the $\dot{\phi}_i^2$ term can be ignored in Eq.~(\ref{eq-GR}).  The corresponding condition can be expressed as ranges of a few combinations of the random coefficients, which makes the probability more straightforward to quantify.\footnote{There are some alternatives for multifield slowroll inflation\cite{DvaKac03,GreHor09,Yan12}, for which one basically trade the range of coefficients for other arrangements.  It becomes less obvious to quantify the probability.  We will stay within the simplest cases and hope it becomes obvious that for the behavior of large number of fields, $N\sim500$, our conclusions are generic.}

Conditions for the standard gradient flow slowroll inflation can be found in some recent works\cite{LidLyt09,Yan12}.  The first slowroll condition requires a small gradient,  
\begin{equation}
\epsilon \equiv \frac{M_P^2 (\vec\nabla V)^2}{2 V^2}\ll1~.
\end{equation}
This is already a strong sign for an exponential suppression.  Since $\vec{\nabla} V$ is an $N$ dimensional vector, it has $N$ components.  If there is no particular correlation between the components, roughly $N$ numbers have to to be simultaneously tuned.  Assuming that $V_0$ is the typical value of $V$ and $\nu$ is the typical value of $|\vec\nabla V|$, the probability to satisfy the first slowroll condition is roughly\footnote{In principle, we should have also scan through values of $V$.  However since it is a scalar, we do not find any strong reason that its distribution exhibit interesting properties at large $N$.  So for simplicity, through out this paper we focus on the tuning of other parameters with the same inflation scale $V_0$.}
\begin{equation}
P_{\rm first} \approx S_{N-1} \int_0^{\frac{\sqrt{2\epsilon}V_0}{M_P}} \frac{v^{N-1}dv}{\nu^N} 
\sim \epsilon^{N/2}\left(\frac{V_0}{\nu M_p}\right)^N ~.
\label{eq-Pfirst}
\end{equation}
The idea is that if $0$ is not a special value and $\nu\gg(\sqrt{2\epsilon}V_0/M_p)$, then the probability that the value of a vector to be within a small ball is roughly the volume of the ball, therefore a small number to the $N$th power.  For the purpose of our estimation, we only keep those small unitless factors which are related to some physical parameters. 

The strong second slowroll condition requires that the projection of second derivatives along the gradient direction is small.
\begin{eqnarray}
\hat{V}_1 &\equiv& \frac{\vec\nabla V}{|\vec\nabla V|}~, \nonumber \\
\stackrel{\leftrightarrow}{V_2} &\equiv& 
\frac{M_P^2(\partial_i\partial_jV)}{V}~, \nonumber \\
\xi &\equiv& \sqrt{\hat{V}_1\cdot\stackrel{\leftrightarrow}{V_2}\cdot\stackrel{\leftrightarrow}{V_2}\cdot\hat{V}_1} \ll1~. 
\label{eq-secondstrong}  
\end{eqnarray}
Here we further require that the classical trajectory is perturbatively stable. Given $\{\lambda_i\}$ as the eigenvalues of $\stackrel{\leftrightarrow}{V_2}$, not only some of the $\lambda_i^2$ needs to be small to guarantee a small projection in Eq.~(\ref{eq-secondstrong}), but also the non-small ones have to be positive.  The probability includes two factors: first the matrix $\stackrel{\leftrightarrow}{V_2}$ needs to be tuned for the above property, then $\hat{V}_1$ needs to sit mostly in the subspace of the small $\lambda_i$'s.  Let $\lambda_0$ be the untuned typical value of $|\lambda_i|$, we have
\begin{eqnarray}
P_{\rm second} &\sim&
\sum_{n=1}^{N} \frac{C^N_n}{2^{N-n}}
\prod_{i=1}^n \left(\int_{-\xi}^\xi \frac{d\lambda_i}{\lambda_0} \int_{-1}^1dv_i\right) 
\delta\left(\cos\theta-\sqrt{\sum_{i=1}^n v_i^2}\right)
\\ & & 
\int_0^{\sin^{-1}\left[\frac{\sqrt{\xi^2-\sum_{i=1}^n \lambda_i^2v_i^2}}{\lambda_0}\right]}
\left(\sin\theta\right)^{N-n-1}d\theta \nonumber \\
&\sim& \sum_{n=1}^{N} \frac{C^N_n}{2^{N-n}}
\left(\frac{2\xi}{\lambda_0}\right)^n 
\left(\frac{\xi}{\lambda_0}\right)^{N-n}
\label{eq-Psecondsum}
\end{eqnarray}
Our rough estimation starts by summing over $n$, the number of eigenvalues $\lambda_i$ which have been tuned small.  Namely, the number of light fields.  For those $(N-n)$ untuned values we only include the $(1/2)$ factor that makes them positive, and later assume that they all take the typical value $\lambda_0$.  We treat the tuning of different $\lambda_i$ as being independent with a flat measure.  The delta function and the complicate integration range comes from the orientation of the unit vector $\hat{V}_1$ constrained by Eq.~(\ref{eq-secondstrong}). $v_i$ stands for the components of $\hat{V}_1$ and $\theta$ is the angle between $\hat{V}_1$ and the $n$ dimensional subspace of the tuned $\lambda_i$.  It is no more than a formality and for our purpose we can roughly simplify it to the last line, where apparently the extra orientation to align $\hat{V}_1$ exactly balances the suppression to tune more $\lambda_i$ to be small.  This of course depends sensitively on our choice of measure for the value of $\lambda_i$ and should not be learned as a general lesson.  However it is clear enough that we will have an overall suppression that is again a small number to the $N$th power. 

Long story short, for multifield slowroll inflation one needs to tune an $N$ dimensional vector and align it with an $N$ dimensional matrix.  Both tunings are naturally suppressed by something to the power $N$ as shown in Eq.~(\ref{eq-Pfirst}) and (\ref{eq-Psecondsum}).

Before proceeding to study the relative probability distribution for the number of e-foldings, we shall make another simplification.  For a random matrix $\stackrel{\leftrightarrow}{V_2}$, the eigenvalues actually do not have independent distributions.  A common behavior is the eigenvalue repulsion\cite{CheShi11,MarMcA11,BatBat12}, such that tuning more than one $\lambda_i$ to be small is even harder than in Eq.~(\ref{eq-Psecondsum}).  So it seems reasonable to keep only the $n=1$ term in Eq.~(\ref{eq-Psecondsum}). 
\begin{eqnarray}
P_{\rm second} &\sim& \int_{-\xi}^\xi \frac{d\lambda}{\lambda_0}
\int_0^{\sin^{-1}\frac{\sqrt{\xi^2-\lambda^2}}{\sqrt{\lambda_0^2-\lambda^2}}}
\cos\theta\left(\sin\theta\right)^{N-2}d\theta 
\label{eq-Psecond} \\ 
\nonumber &\sim&
\int_{-\xi}^\xi \frac{d\lambda}{\lambda_0}
\int_0^{\sin^{-1}\frac{\sqrt{\xi^2-\lambda^2}}{\lambda_0}}
\left(\sin\theta\right)^{N-2}d\theta
\end{eqnarray}
Also, making the other $(N-1)$ eigenvalues to be all positive will be harder than just $2^{-(N-1)}$.  It might provide another suppression factor that depends even more strongly on $N$, which goes as $e^{-a(N-1)^2}$.  It is a common factor that does not affect the relative probability distributions for $N_e$, so we will not include it here.  But later we will need to consider it when checking for pathologies in Sec.\ref{sec-check}.  

Note that by this choice, the resulting slowroll model will be effectively single-field.  We are just keeping track of the tunings needed to embed it in a multifield background.  In this effectively single field situation, $\xi$ will be roughly equivalent to the standard second slowroll parameter $\eta$.  However we will not enforce the observational constraints on $\epsilon$ or $\xi$.  Since in principle this particular point can be anywhere on an inflationary trajectory, not necessarily within our observable window.

\subsection{Power Law Suppression for More E-Foldings}
\label{sec-efold}

From the previous section, all the required tuning seems to acquire the power $N$.  Our next step is to include the dependence on $N_e$, and determine whether or not such dependence also acquires the power $N$. 

Assume that $\phi_1$ is the direction of $\vec\nabla V$ and $\lambda_1'$ is tuned small\footnote{The direction $1$ and $1'$ are not necessary identical, but must be very close as given by the range of $\theta$ in Eq.~(\ref{eq-Psecond})}.  Along this direction, since both the first order and second order terms are small, the third order term becomes relevant.  Without loss of generality, we assume $\phi_1=0$ at this point and expand the potential along $\vec\nabla V$.
\begin{equation}
V = V_0 + c_1\phi_1 + c_2\phi_1^2 +c_3\phi_1^3~.
\end{equation}
The total number of e-foldings supported by this potential is roughly
\begin{equation}
N_e \sim \frac{V_0}{M_p^2}(3c_1c_3-c_2^2)^{-1/2}~.
\end{equation}
When $c_2^2>3c_1c_3$, a local minimum emerges and traps the fields.  That situation is automatically excluded from our calculation.  We can also choose both $c_1$ and $c_3$ to be positive without loss of generality.  This number of e-foldings occurs within the field range
\begin{equation}
\Delta\phi_1 \sim \frac{2\sqrt{3c_1c_3-c_2^2}}{3c_3}~, 
\label{eq-range}
\end{equation}
and centered at
\begin{equation}
\delta\phi_1 \sim -\frac{c_2}{3c_3}~.
\label{eq-shift}
\end{equation}
With these choices, $\phi_1=0$ is roughly the starting point of an inflation trajectory from which the field slowly rolls down toward the $-\phi_1$ direction.  The probability distribution for $N_e$ is given by the combination of Eq.~(\ref{eq-Pfirst}),~(\ref{eq-Psecond}), an integral over the untuned parameter $c_3$, and a delta function of $N_e$.
\begin{eqnarray}
P(N_e) &=&
\int_0^{\frac{\sqrt{2\epsilon}V_0}{M_P}} \frac{v^{N-1}dv}{\nu^N} 
\int_{-\xi}^\xi \frac{d\lambda}{\lambda_0}
\int_0^{\sin^{-1}\frac{\sqrt{\xi^2-\lambda^2}}{\lambda_0}}
\left(\sin\theta\right)^{N-2}d\theta \nonumber \\
& & \int \frac{dc_3}{\bar{c}_3} ~ 
\delta\left[N_e - \frac{V_0}{M_p^2}(3c_1c_3-c_2^2)^{-1/2}\right]~.
\label{eq-collect1}
\end{eqnarray}
Here again, we pick the measure for $c_3$ to be flat for no better reason otherwise.  In order to keep the entire expression unitless, we include $\bar{c}_3$ as the typical value of $c_3$.  Integrating over the delta function turns $c_3$ into a function of $N_e$, $c_1$ and $c_2$.  We then perform the other two integrals with the following substitutions:
\begin{eqnarray}
c_1 &=& v~, \\
c_2 &=& \frac{V_0}{2M_p^2}(\lambda\cos^2\theta+\lambda_0\sin^2\theta)
\approx \frac{V_0}{2M_p^2}\lambda~.
\end{eqnarray}  
As before, keeping only ratios of physical parameters, we have
\begin{equation}
P(N_e)\sim \left(\frac{\sqrt{\epsilon}V_0}{M_p\nu}\right)^{N-1}
\left(\frac{\xi}{\lambda_0}\right)^N
\left(\frac{V_0^2}{\nu\bar{c}_3 M_p^4N_e^3}\right)~.
\label{eq-Prough}
\end{equation}
We carefully arrange the factors into three brackets, coming separately from $P_{\rm first}$, $P_{\rm second}$, and the integral of $c_3$ with the delta function.  Only the last one depends on $N_e$.  

At the first glance, it should be quite surprising that the dependence on $N_e$ does not care about the number of fields.  It becomes more transparent after thinking about the corresponding physical interpretation.  The first two tunings are to embed a single field inflation in a multifield theory.  They involve $N$ dimensional vectors and matrices, so they introduce something to the $N$th power.  After those two quantities are tuned, we have already picked a direction.  Tuning for more e-foldings essentially concerns this particular direction only, therefore no $N$th power involved.

Although this result is the same as in \cite{AgaBea11}, through this interpretation we realize that it is still premature to conclude.  The above calculation only considers the cases where inflation ends in the standard single-field manner.  In other word we implicitly assumed that slowroll inflation can take advantage of the entire field range given by Eq.~(\ref{eq-range}) and (\ref{eq-shift}).  That is of course too optimistic.  When this model is embedded in an $N$ dimensional field space, there are other ways for inflation to end.  A slightly different but related perspective: $\phi_1^3$ is not the only third order term.  At the starting point it is the only relevant one, but other terms may become important as we move along the inflation trajectory.  For example, consider
\begin{equation}
V =  V_0 + c_1\phi_1 + c_2\phi_1^2 +c_3\phi_1^3 + 
\frac{\lambda_0 V_0}{M_p^2} \phi_2^2 + q \phi_1 \phi_2^2~.
\label{eq-thirdcross}
\end{equation} 
At 
\begin{equation}
\phi_1 = \frac{-\lambda_0 V_0}{q M_p^2}~,
\end{equation}
the $\phi_2$ direction is destabilized.  We will not have a stable slowroll solution beyond this point even though it might be still within the combined range of Eq.~(\ref{eq-range}) and (\ref{eq-shift}).  In other words, multifield inflation can end in ways not captured by the effective single field model\footnote{With untuned parameters, the $\phi_2$ direction quickly becomes very tachyonic and will not support a second stage of hybrid inflation.}.  

We will not dive into detail mechanisms of how inflation ends in these manners.  It is natural to assume that various ways for these other fields to end inflation have nothing to do with the tuned parameters in the $\phi_1$ direction.  We will simply model these possibilities as a $\Delta\phi_{\rm max}$.  Namely, even if we start at a point which is tuned to have effectively single field inflation, after moving along the slowroll trajectory $\Delta\phi_{\rm max}$ distance away, inflation is no longer supported.  Practically, this means that Eq.~(\ref{eq-collect1}) should include a step function.
\begin{eqnarray}
\label{eq-collect2}
P(N_e) &=&
\int_0^{\frac{\sqrt{2\epsilon}V_0}{M_P}} \frac{v^{N-1}dv}{\nu^N} 
\int_{-\xi}^\xi \frac{d\lambda}{\lambda_0}
\int_0^{\sin^{-1}\frac{\sqrt{\xi^2-\lambda^2}}{\lambda_0}}
\left(\sin\theta\right)^{N-2}d\theta  \\
& & \int dc_3 ~ 
\delta\left[N_e - \frac{V_0}{M_p^2}(3c_1c_3-c_2^2)^{-1/2}\right]~\nonumber
\Theta\left(\Delta\phi_{\rm max}-|\delta\phi_1|-\frac{\Delta\phi_1}{2}\right)~.
\end{eqnarray}
So, there is a chance that this bound on $\Delta\phi$ forces us to further tune the $N$ dimensional vector and matrix for more e-foldings.  For example one may imagine that longer inflation requires a larger field space distance, thus having a higher risk of running into $\Delta\phi_{\rm max}$.

To evaluate Eq.~(\ref{eq-collect2}), we will again eliminate $c_3$ with the delta function.  The step takes place at
\begin{equation}
\Delta\phi_{\rm max}=
|\delta\phi_1|+\frac{\Delta\phi_1}{2}=
\frac{c_1}{c_2^2+\frac{V_0^2}{N_e^2M_p^4}}
\left(|c_2|+\frac{V_0}{N_eM_p^2}\right)~.
\label{eq-rangeBound}
\end{equation}
The only way to modify the single field conclusion is when the theta function replaces some of the integration limit.  When 
\begin{equation}
\Delta\phi_{\rm max}<2M_p \frac{\sqrt{2\epsilon}}{\xi}~,
\end{equation}
the replacement always occurs and gives us
\begin{equation}
P(N_e) \approx \left(\frac{V_0^2}{\nu\bar{c}_3 M_p^4N_e^3}\right) 
\int_0^{\frac{\Delta\phi_{\rm max}V_0}{2M_p^2}
\frac{\lambda^2+4N_e^{-2}}{|\lambda|+4N_e^{-1}}} 
\frac{v^{N-2}dv}{\nu^{N-1}} 
\int_{-\xi}^{\xi} 
\frac{(\xi^2-\lambda^2)^{\frac{N-1}{2}}d\lambda}{\lambda_0^N}~.
\label{eq-NdNe}
\end{equation}  

One can just evaluate this integral.  For our purpose, it is more intuitive to analyze it by splitting into two regimes.  When $|\lambda|>2N_e^{-1}$, the RHS of Eq.~(\ref{eq-rangeBound}) is roughly $\frac{c_1}{|c_2|}$; when $|\lambda|<2N_e^{-1}$, it is roughly $\frac{c_1N_eM_p^2}{V_0}$.  Thus we have two possible behavior relatively for long or short inflations,
\begin{eqnarray}
P(N_e) &\approx& \left(\frac{V_0^2}{\nu\bar{c}_3 M_p^4N_e^3}\right)
\left(P_s + P_l\right)~, \\
P_s &=& 
\int_0^{\frac{\Delta\phi_{\rm max}V_0}{N_eM_p^2}} \frac{v^{N-2}dv}{\nu^{N-1}} 
\int_{-N_e^{-1}}^{N_e^{-1}} 
\frac{(\xi^2-\lambda^2)^{\frac{N-1}{2}}d\lambda}{\lambda_0^N}~, \\ 
P_l &=& 
2\int_0^{\frac{\Delta\phi_{\rm max}V_0\lambda}{2M_p^2}} \frac{v^{N-2}dv}{\nu^{N-1}} 
\int_{N_e^{-1}}^\xi \frac{(\xi^2-\lambda^2)^{\frac{N-1}{2}}d\lambda}{\lambda_0^N}~.
\end{eqnarray} 

When $N_e\lesssim\xi^{-1}$, the integration range for $P_l$ does not exist so we indeed get a suppression of $N_e^{-N}$ from $P_s$.  However, $P_l$ dominates for when $N_e\gtrsim\xi^{-1}$ and it has no extra $N_e$ dependent suppression.  The exact behavior of Eq.~(\ref{eq-rangeBound}) is not smooth between the two regimes and the choice of $\xi$ is somewhat arbitrary.  So we shall not take the detail behavior of Eq.~(\ref{eq-NdNe}) too seriously.  However, the qualitative conclusion is solid.  Apparently, tuning for a large number of e-foldings is not constrained by the field range bound $\Delta\phi_{\rm max}$\footnote{Classically, the potential can be tuned to provide infinite e-foldings in a finite field range.  Including quantum fluctuations, it would have entered eternal slowroll inflation instead.  Our argument still works in that case.}.  Therefore, if the multifield endings of inflation are characterized by $\Delta\phi_{\rm max}$, then in the large $N_e$ limit, longer inflation is still only suppressed by a mild power law, $N_e^{-\alpha}$ with $\alpha\sim3$ just like in a single field model.

\subsection{Attraction Region}
\label{sec-attract}

Finally, we shall include the volume factor from the attraction region.  First we briefly review how it works in single field open inflation\cite{FreKle05}. Let $0>\phi>-\Delta\phi$ be the inflation region.  The number of e-folding is $n_e$ if the field starts at $\phi=0$ and runs through the entire region.  If one starts at $-\delta\phi$ instead, the solution will inflate for roughly $n_e\frac{\Delta\phi-\delta\phi}{\Delta\phi}$ e-foldings.  The interesting behavior for open inflation is that if one starts at $\delta\phi$, where the slope of potential is very steep, it will only overshoot up to $-\delta\phi$ and also inflate for $n_e\frac{\Delta\phi-\delta\phi}{\Delta\phi}$ e-foldings.  Thus the weighting from the attractor region works as the following.
\begin{equation}
P_{\rm weighted}(N_e) = \int_{N_e}^{\infty} dn_e~
\int_{-\Delta\phi}^{\Delta\phi}d\delta\phi~P(n_e) ~
\delta\left(N_e - n_e\frac{\Delta\phi-|\delta\phi|}{\Delta\phi}\right)~.
\end{equation}
In other words, we are not calculating the field space volume of a single region.  A Region that can support exactly $N_e$ efoldings means the field has to start at exactly one correct point.  Regions supporting more efoldings also contribute only $N_e$ if starting at a particular shell (in the single field case, 2 points).  So we get an effective ``volume'' by integrating over the shells from different regions.

For multifield models, we shall first dispel a na\"ive picture that is somewhat misleading.  For example, with 2 fields, some might take the typical potential as
\begin{equation}
V = V_{\rm SR}(\phi_1) + \frac{\lambda_0}{2M_p^2}\phi_2^2~.
\end{equation}
If this is the case, $\phi_2$ undergoes a damped oscillation which has almost no effect on the inflationary motion of $\phi_1$.  In that case, the attraction region will be exponentially large in the $\phi_2$ (in general, orthogonal) directions.  This may overcompensate the rareness of slowroll regions and jeopardize the validity of our entire analysis.

In the above situation, $\phi_1$ does not pick up a any extra velocity from a large displacement in $\phi_2$.  It clearly should not be taken as the typical situation.  Recall that the third order cross terms like $\phi_1\phi_2^2$ are not suppressed, so something like Eq.~(\ref{eq-thirdcross}) represents our situation more faithfully.  It implies that the mass of $\phi_2$ changes with $\phi_1$, and also the slope of $\phi_1$ changes with $\phi_2$.  When the amplitude of $\phi_2$ oscillation is still large, most likely $\phi_1$ is not slowly rolling.  Depending on the signs of these third order terms, either certain field space distance is squandered during the damped oscillation, or the field does not even roll back to this inflation trajectory.  

Again we shall not dive into various details about possible field motions in $N$ dimensional space.  Our first approximation is to treat all orthogonal directions equally, which is true in a statistical sense.  So what matters is the total magnitude
\begin{equation}
\delta\phi_{\rm ort} = \sqrt{\sum_{i=2}^N \phi_i^2}~.
\end{equation}
Then we simply define the quantity $\delta\phi(\delta\phi_1,\delta\phi_{\rm ort})$.  This refers to the amount of field displacement along the inflation trajectory that was not spent in slowroll motion.  We already know from the single field example that
\begin{equation}
\delta\phi(\delta\phi_1,0) = |\delta\phi_1|~. 
\end{equation}
For small $\delta\phi_{\rm ort}$, we can assume
\begin{equation}
\delta\phi(0,\delta\phi_{\rm ort}) = c|\phi_{\rm ort}|^\beta~,
\end{equation}
for some appropriate $c>0$ and $\beta>0$.\footnote{For example, Eq.~(\ref{eq-thirdcross}) with $q>0$ leads to $\beta=2$.}  We will ignore the possible cross dependence and assume the following particular form for small $\delta\phi$.
\begin{equation}
\delta\phi(\delta\phi_1,\delta\phi_{\rm ort})=
\sqrt{(\delta\phi_1)^2+c^2(\delta\phi_{\rm ort})^{2\beta}}~.
\label{eq-squander}
\end{equation}
The exact form does not really matter.  The constant $\delta\phi$ surface simply provides the $(N-1)$ dimensional shell contribution from a region that we can integrate over.

By analogy to the single field calculation, we have
\begin{eqnarray}
P_{\rm weighted}(N_e) &=& \int_{N_e}^\infty dn_e ~
\int d\delta\phi_1 ~ 
\int (\delta\phi_{\rm ort})^{N-2} d\delta\phi_{\rm ort}~ 
\nonumber \\
& & P(n_e)~\delta\left(N_e - n_e\frac{\Delta\phi-
\delta\phi(\delta\phi_1,\delta\phi_{\rm ort})}{\Delta\phi}\right)~.
\end{eqnarray}

Eq.~(\ref{eq-squander}) allows us to change variable to $\delta\phi$ to get
\begin{eqnarray}
P_{\rm weighted}(N_e) &\sim& \int_{N_e}^\infty dn_e ~
\int (\delta\phi)^{\frac{(N-1)}{\beta}} d\delta\phi~P(n_e) ~
\delta\left(N_e - n_e\frac{\Delta\phi-\delta\phi}{\Delta\phi}\right)~
\nonumber \\
&\sim& \int_{N_e}^\infty dn_e ~ n_e^{-4}
\left(1-\frac{N_e}{n_e}\right)^{\frac{(N-1)}{\beta}} \nonumber \\
&=& N_e^{-3} \int_0^1 (1-x)^2 x^{\frac{(N-1)}{\beta}} dx~.
\end{eqnarray}
So, we see that the attraction mechanism does not change the $N_e$ dependence.  This is again because the attraction mechanism is naturally described by the field space distance, $\delta\phi$.  We already learned from the previous section that tuning for more e-foldings is quite parallel to the field space distances, so there is little reason to care.

\section{Checking for Pathologies}
\label{sec-check}

In the multiverse scenario, one repeatedly runs into situations that certain aspect of our universe seems rare.  One should not be scared and prematurely conclude that the multiverse is wrong.  For every trait of rareness, one can make specific predictions in the form of relative probabilities, and check if such predictions are in conflict with experiments or observations.\footnote{Or one could try to see if the rareness makes it impossible to realize certain necessary condition for our universe in the entire landscape.  Given the exponentially large size of the landscape, those efforts have been inconclusive.}

For example, the famous Boltzmann Brain (BB) problem is actually the following relative probability.
\begin{equation}
\frac{P({\rm consistent \ evolution}~|~{\rm current \ observation})}
{P({\rm random \ outcome}~|~{\rm current \ observation})}
=\frac{P_{OO}}{P_{BB}}~.
\end{equation}
Given the current state of our brains that observes our surroundings, one can try to predict how the world looks like, say, one minute in the future.  An ordinary observer (OO) would see that everything still evolves according to the known physical laws, while a Boltzmann Brain would ``think'' that it is seeing totally random outcome, or simply itself will dissipate.

So, for any theory that predicts $\frac{P_{BB}}{P_{OO}}\gg1$, it constantly runs into contradictions with observations (every minutes per observer in our example).  It is ruled out by an exponentially high confidence level in any practical standard.

It has been shown that in most of the successful measures,
\begin{equation}
\frac{P_{BB}}{P_{OO}}=\frac{\Gamma_{BB}}{\Gamma_{decay}}~.
\end{equation}
Thus if all the BB habitable vacua decay before producing them, the above ratio is much less than one.

However, all existing analysis assumed that the number of BB friendly vacua is comparable to those producing OO.  The production of OO requires slowroll inflation, which we have shown to be exponentially rare.  So one should include some more suppression factors,
\begin{equation}
\frac{P_{BB}}{P_{OO}}\sim\frac{\Gamma_{BB}}
{\Gamma_{decay}\epsilon^{\frac{N-1}{2}}\xi^Ne^{-aN^2}}~.
\end{equation}
Here $\epsilon$ and $\xi$ are the slowroll parameters coming from Eq.~(\ref{eq-Prough}); $e^{-aN^2}$ is the possible suppression factor coming from eigenvalue repulsion where $a$ is an order one number. 

A conservative estimation from \cite{Pag06b} gives $\Gamma_{BB}\sim \exp[-10^{42}]$~.  So even if we take $\epsilon$ and $\xi$ as small as the observation bound and $N\sim500$, this factor is obviously not enough to revive the Boltzmann Brain problem.

Another relative probability we should check is
\begin{equation}
\frac{P({\rm detect \ open \ curvature}~|~{\rm current \ curvature \ bound})}
{P({\rm improve \ curvature \ bound}~|~{\rm current \ curvature \ bound})}
=\frac{\int_{n_1}^{n_2} P(N_e) dN_e}{\int_{n_2}^{\infty} P(N_e) dN_e}~.
\end{equation} 
If the probability of having more e-foldings is significantly suppressed, then people sitting with the data from WMAP1\cite{Spe03} should expect to be on the verge of seeing a non-zero curvature, instead of large improvement of the bound consistent with zero.  It has already be shown in \cite{FreKle05} that a landscape of single field inflation has $P(N_e)\propto N_e^{-4}$, which is a mild enough suppression to avoid such problem.  Our analysis shows that for a multifield landscape such conclusion is still true.  Imagine if we had instead shown that $P(N_e)\propto N_e^{-N}$, then the improvement of curvature bound from requiring $n_1=30$ e-foldings to $n_1=50$ e-foldings would have had a probability of about $(30/50)^{500}$.  It would have been a serious contradiction with observations. 

\section{Discussion}
\label{sec-dis}
We provided a heuristic argument that in a multifield landscape, the FVEI framework provides the following probability distribution to realize slowroll inflation.
\begin{equation}
P(N_e) = A N_e^{-\alpha}~,
\end{equation}
where $A$ is exponentially small and $\alpha\sim3$.  Although we focused on one particular type of slowroll model to write down specific equations, we expect this behavior to be generic.  Basically, the number of conditions to be tuned for slowroll grows with the number of fields.  That is why $A$ is an exponentially small number and the exponent depends on $N$.  Longer e-foldings only requires tuning in one particular direction and can be produced in a confined region in the field space, so $\alpha$ does not grow with $N$.  We also argued that such probability is still consistent with the multiverse scenario.  

The next interesting question is, can the multiverse scenario tell us that among so many proposed slowroll models, which ones are more likely to make our universe, thus deserve more attention.  We should first remind the readers that if the multiverse selection rule assigns a $(90\%,10\%)$ probabilities to two models, that is pretty useless.  Since we only have one universe to observe, being as rare as a few sigma event is not a sharp contradiction.  A useful selection rule needs to give exponential relative probabilities, like the ones we checked in Sec.\ref{sec-check}.  Thus, the fact that $A$ is exponentially small is actually cruicial.  Since the ratio of exponentially small numbers are typically exponential, this suggests the possibility of useful selection rules.

Obviously, within the scope of this paper we cannot provide such a specific rule.  Even the measures of the parameters in our analysis are quite arbitrary.  However, there is a very direct way to make predictions.  We have assumed that the tunneling ends in a random place, and it does not correlate with the local properties of $V$ that controls whether we can have slowroll inflation.  If in some models, the tunneling paths always end at places supporting slowroll inflation, then such model does not suffer from the exponential suppression.

We have actually took a quite tortuous way to demonstrate such a simple idea.  In the FVEI picture slowroll inflation comes after a tunneling, so it prefers a slowroll model that such sequence is likely.  If someone can recognize regions on the landscape that tunneling paths must end, then one should focus on slowroll models supported by such regions.  Due to the current growth of interests on multifield tunneling paths\cite{Yan09,AguJoh09a,BroDah10,AhlGre10,BroDah11}, this goal may come within our reach soon.  Certain global properties of the effective potential enforce a detour to a special direction.  There are currently two examples for such excursion.  One goes toward large compactification volume\cite{Yan09,AguJoh09a,BroDah10,BroDah11}, and the other goes toward a strongly warped throat\cite{AhlGre10}.  If one can design a slowroll model that connects with the tunneling path returning from these special directions, they will not suffer from the exponential suppression and may be the most likely slowroll model from the multiverse point of view.

\acknowledgments 
We thank Raphael Bousso and Ben Freigovel for stimulating discussions.  I-Sheng Yang thanks University of Amsterdam, Lorentz Center and National Taiwan University for their hospitality while completing this work.  This work is supported in part by the US Department of Energy, grant number DE-FG02-11ER41743.

\bibliographystyle{utcaps}
\bibliography{all}

\end{document}